\documentclass[pra,twocolumn,superscriptaddress,amsmath,amssymb,showpacs]{revtex4-1}

\usepackage{color}
\usepackage{dcolumn,amsmath,amssymb}

\usepackage[caption=false]{subfig}
\usepackage{graphicx}

\begin{document}

\title{ Pendular trapping conditions for ultracold polar molecules enforced by external electric fields}

\author{Ming Li$^a$, Alexander Petrov$^{a,b}$, Constantinos Makrides$^{a,c}$, Eite Tiesinga$^c$ and Svetlana Kotochigova$^{a\ast}$\\
             \thanks{$^\ast$Corresponding author. Email: skotoch@temple.edu} \\
          $^{a}${\em Department of Physics, Temple University, Philadelphia, PA 19122-6082, USA}\\
          $^{b}${\em St. Petersburg Nuclear Physics Institute, Gatchina, 188300; Division of
                     Quantum Mechanics, St. Petersburg State University, 198904, Russia}\\
		     $^{c}${\em Joint Quantum Institute and Joint Center for Quantum Information and Computer Science, 
		     National Institute of Standards and Technology and University of Maryland, Gaithersburg MD 20899, USA}     }

\begin{abstract}
We theoretically investigate trapping conditions for ultracold polar
molecules in optical lattices, when  external magnetic and electric fields
are simultaneously applied.  Our results are based on an accurate
electronic-structure calculation of the polar $^{23}$Na$^{40}$K polar
molecule in its absolute ground state combined with a calculation of its
rovibrational-hyperfine motion.  We find that an electric field strength of
$5.26(15)$ kV/cm and an angle of $54.7^\circ$ between this field and the
polarization of the optical laser lead to a trapping design for
$^{23}$Na$^{40}$K molecules where decoherences due laser-intensity
fluctuations and fluctuations in the direction of its polarization are kept
to a minimum. One standard deviation systematic and statistical
uncertainties are given in parenthesis.  Under such conditions pairs of
hyperfine-rotational states of $v=0$  molecules, used to induce tunable
dipole-dipole interactions between them, experience ultrastable, matching
trapping forces.  
\end{abstract}

\date{\today}


\maketitle

\section{Introduction}
The successful creation of near quantum-degenerate gases of polar molecules in their absolute rovibrational 
ground state (\textit{e.g.} KRb~\cite{KNi2008}, 
 RbCs~\cite{Takekoshi2014,Cornish2014},
NaK~\cite{ParkPRL2015}, and NaRb~\cite{Wang2016}),
opened up the possibility of studying controlled collective 
phenomena, ultracold chemistry, quantum computing, and of performing 
precision measurements with polar molecules.  
In most of these applications polar molecules likely need to be held in periodic, optical potentials 
induced by external laser fields, where two or more of their 
rotational hyperfine states are manipulated and accurate measurement 
of the transition frequency between these levels is required. 

Dynamic Stark shifts of these hyperfine levels in the presence of 
trapping laser fields are generally different, depending  
on a range of experimental parameters. As a result the system is 
sensitive to laser-intensity fluctuations leading to 
uncertainties in the transition-energy measurements or decoherence
when attempting to couple the states of interest for quantum control.
Thus, a careful selection of trapping conditions, where a pair of internal 
states experience identical trapping potentials, can bring 
substantial benefits. Such experimental conditions are called {\it magic}.

{\it Magic} electric-field values for polar molecules have applications
in the realm of many-body, non-equilibrium spin physics.   This
includes  samples of molecules with long-range dipole-dipole
interactions tailored by static electric fields 
or by a combination of electric  and resonant microwave fields
\cite{Zoller2006,Zoller2007,ZollerPRL2007}.  Working at a {\it magic}
electric field, for example,  ensures that spatial laser-intensity
inhomogeneities across a large sample do not significantly change
the resonant condition for the microwave  field.  Initial experimental
realizations applied  electric fields  up to a few kV/cm. Larger
electric field apparatuses are now under development
\cite{Gempel2016,JCovey2016}  with fields above 10 kV/cm promising
larger dipole moments and individual addressing and detection.

In a previous study \cite{Kotochigova2006}, we calculated the dynamic
or AC polarizability of polar KRb and RbCs molecules. We located
optical frequency windows, where light-induced decoherence is small,
and determined van-der-Waals potentials between the
molecules~\cite{Kotochigova2010a}.  We matched the AC polarizability
of the $N=0$ and $N=1$ rotational states of these molecules  with
a {\it magic} electric field and  angle between laser polarization and
electric field direction in Ref.~\cite{Kotochigova2010}.  In parallel,
optimal trapping conditions for  homonuclear Rb$_2$ and Cs$_2$ were
studied in Refs.~\cite{Dulieu2011,Dulieu2015}.

For the KRb molecule we extended our calculations  by including
hyperfine coupling between rotation and the nuclear electric
quadrupole moment and found in good agreement with
experiment~\cite{BNeyenhuis2012}.  The coherence time for a rotational
superposition was maximized  at the {\it magic} angle.  Recently,
Ref.~\cite{Moses2017} suggested that the coherence time is now
limited by  laser-intensity fluctuations across the molecular sample.
Finally, in Ref.~\cite{APetrov2013} we performed an investigation
for rovibronic states of  $^{40}$K$^{87}$Rb when three external
fields are present, i.e. magnetic, electric and trapping-laser
fields.  The magnetic field was relatively large with a strength
near 50 mT and  the hyperfine coupling between the nuclear spins
and other angular momenta had a negligible effect.

In the current study we propose alternative means to extend coherence
times for superpositions of molecular rotational states. We focus
on decreasing the dependence of the dynamic polarizability on the
light intensity and fluctuations or, equivalently, on minimizing
the hyperpolarizability with respect to intensity by orienting polar
molecules in a strong uniform electric field and creating so called
{\it pendular} rotational states
\cite{HHughes1947,JRost1992,HLoesch1995,BFriedrick2008}.  In such
arrangement the DC Stark effect  dominates  and the complex coupling
between hyperfine states with different Stark shifts goes to zero.

We present a theoretical study of the dynamic polarizability of
rotational hyperfine states of ultracold NaK molecules. The NaK
molecule has a large permanent electric dipole moment and is chemically stable against atom-exchange
reactions \cite{WSebastian2016}.  A long-lived quantum gas of
fermionic $^{23}$Na$^{40}$K molecules was created in its absolute
ground state using a magnetic field of 8.57 mT \cite{JPark2015}.
Each of its rotational states $|N, m\rangle$ has 36 hyperfine
states.  At this magnetic field the hyperfine coupling between
nuclear spins and orbital angular momenta is strong and combined
fluctuations in the magnetic and electric field and trapping laser
can induce drastic changes in the complex hyperfine structure. On
the other hand, due to the large dipole moment of polar molecules,
our static electric field will force a simplification of  the
hyperfine structure.

The paper is set up as follows. In Sec.~\ref{sec:th}, we present
the  molecular Hamiltonian for ground-state alkali-metal
dimers and the pendular model for strong electric fields. In
Sec.~\ref{sec:NaK}, we apply our theory to non-reactive $^{23}$Na$^{40}$K
to elucidate the role of an electric field, and give 
its {\it magic} trapping conditions. We summarize in Sec.~\ref{sec:sum}.

\section{Theory}
\label{sec:th}

The effectiveness of trapping ultracold polar alkali-metal molecules
with optical lasers is determined by the (real) dynamic polarizability
of their ro-vibrational-hyperfine states. The polarizability of a molecular
eigenstate $i$ with energy ${\cal E}_i$ under the influence of a
linearly-polarized laser with frequency $\omega$ and intensity
$I_\mathrm{trap}$ is defined as the derivative $\alpha_{\mathrm{dyn},i}
= -d{\cal E}_i/dI_\mathrm{trap}$. The dynamic polarizability can
then studied as a function of the strength and orientation of static
magnetic and electric fields.  Eigenenergies of hyperfine states
need to be calculated with care.  The  starting point is an effective
molecular Hamiltonian that contains all {\it internal} and {\it external}
interactions. It is described in subsection~\ref{sec:ham}.

With even a moderate electric field, the DC Stark effect together
with the rotational energy dominate over other interactions. A
simplified, pendular model is then sufficient. It is given in subsection
\ref{sec:simp} and will provide  physical insight as well as an
easy way to calculate the total polarizability in this regime.

\subsection{Molecular Hamiltonian}
\label{sec:ham}

We focus on rotational, hyperfine states of the lowest vibrational
level of the ground singlet X$^1\Sigma^+$ electronic potential of
alkali-metal dimers in the presence of a magnetic and electric field
as well as a trapping laser. Our notation and conventions for angular 
momentum algebra are based on Ref.~\cite{Brink1993}.
The effective Hamiltonian
is~\cite{JAldegunde2008,APetrov2013}
\begin{equation}
H = H_\mathrm{rot} + H_\mathrm{hf} + H_\mathrm{Z}
    + H_\mathrm{E} + H_\mathrm{pol} \,, 
\label{eq:ham}
\end{equation}
where
\begin{gather}
   H_\mathrm{rot} = B_{v=0}\boldsymbol{N}^2 \, , \quad \quad\quad
   H_\mathrm{hf} = \sum_{k=a,b}\boldsymbol{V}_k\cdot\boldsymbol{Q}_k \,, \nonumber\\
   H_\mathrm{Z} = -\sum_{k=a,b} g_k\mu_B \,
                   \boldsymbol{I}_{k}\cdot\boldsymbol{B} \,, \quad\quad 
   H_\mathrm{E} = -\boldsymbol{d}\cdot\boldsymbol{E} \,, \nonumber
   \end{gather}
 and 
 \begin{eqnarray*}
	H_\mathrm{pol} &=& -\frac{1}{3} [\alpha_{||}(\omega)+2\alpha_\perp(\omega)]I_\mathrm{trap} \\
	&&
      \quad - \frac{\sqrt{6}}{3}[\alpha_{||}(\omega)-\alpha_\perp(\omega)] 
      T_2(\boldsymbol{\hat \epsilon},\boldsymbol{\hat \epsilon)}\cdot C_2(\alpha,\beta) I_\mathrm{trap} \,. 
       \end{eqnarray*}
The rotational  $H_\mathrm{rot}$ and hyperfine $H_\mathrm{hf}$
Hamiltonian describe the {\it internal} field-free molecular interactions.
Here, $\boldsymbol{N}$ is the rotational angular momentum operator
of the molecule and $B_{v=0}$ is the rotational constant of vibrational
state $v=0$.  The hyperfine Hamiltonian is  the nuclear
electric-quadrupole interaction, where $\boldsymbol{Q}_k$ is the
electric quadrupole moment of nucleus $k=a$ or $b$ and $\boldsymbol{V}_k$
is the electric field gradient generated by the electrons at the position of that 
nucleus. For nuclear spins $\boldsymbol{I}_k$  with quantum number $I_k>1$ this
interaction is equivalent to
\begin{equation}
   H_\mathrm{hf}=\sum_{k=a,b} \frac{(eqQ)_k}{I_k(I_k-1)} 
           T_2(\boldsymbol{I}_k,\boldsymbol{I}_k) \cdot C_2(\alpha,\beta)\,,
\end{equation}
where $C_{lm}(\alpha,\beta)=\sqrt{4\pi/(2l+1)}Y_{lm}(\alpha,\beta)$,
$Y_{lm}(\alpha,\beta)$ is a spherical harmonic of rank $l$, Euler angles
$\alpha$, $\beta$ and $\gamma$ describe the orientation of the interatomic
axis in a space-fixed coordinate frame, $T_{2m}(\boldsymbol{I}_k,\boldsymbol{I}_k)$ is the rank-2
spherical tensor constructed from nuclear spin operators and $(eqQ)_k$ is the
nuclear electric-quadrupole coupling constant.

The effects of the static magnetic field, the static electric field,
and the trapping laser field are included through the {\it external}
nuclear Zeeman Hamiltonian $H_\mathrm{Z}$, the DC Stark effect
$H_\mathrm{E}$, and  molecule-laser interaction $H_\mathrm{pol}$,
respectively.  In $H_\mathrm{Z}$, $g_k$ is the gyromagnetic ratio
of nucleus $k$, $\boldsymbol{B}$ is the magnetic field, and $\mu_B$
is the Bohr magneton. In $H_\mathrm{E}$, the operator $\boldsymbol{d}$
is the vibrationally-averaged molecular dipole moment  and
$\boldsymbol{E}$ is the static electric field.  The Hamiltonian
$H_{\rm{pol}}$ depends on the frequency-dependent vibrationally-averaged parallel and
perpendicular polarizabilities $\alpha_{\parallel}(\omega)$ and $
\alpha_{\perp}(\omega)$, and laser intensity $I_\mathrm{trap}$.
The two rank-2 tensor operators  capture its dependence on
(linear) laser polarization $\boldsymbol{\hat{\epsilon}}$ and
rotational state of the molecule~\cite{Kotochigova2010}.  (The $\alpha_{\parallel}(\omega)$ and $
\alpha_{\perp}(\omega)$ will be further discussed in Sec.~\ref{sec:pol}.) We neglect contributions
from centrifugal distortions, the rotational Zeeman interaction,
and other hyperfine terms.

\subsection{Basis set, coordinate system, and quantization axis}

It is convenient to find the eigenstates of Eq.~\ref{eq:ham} using 
the uncoupled molecular hyperfine states
\begin{align}
|N,&m,m_a,m_b\rangle\equiv \nonumber\\
 &  \phi_{v=0}(r)\, Y_{Nm}(\alpha,\beta) | \Lambda \rangle   
                  \,| I_a,m_a\rangle | I_b,m_b\rangle, 
\label{eq:channel}
\end{align}
where $\phi_{v=0}(r)Y_{Nm}(\alpha,\beta)$ is the $v=0$ radial
vibrational and rotational wavefunction as a function of the
internuclear separation and orientation $\vec r=(r,\alpha, \beta)$
in spherical coordinates. The function $\phi_{v=0}(r)$ is to good
approximation independent of the rotational quantum number $N$ when
$N$ is small.  The kets $|\Lambda\rangle$ and $| I_k,m_k\rangle$
describe the electronic and nuclear-spin wavefunctions, respectively.

The projection quantum numbers and angles are defined with respect
to a coordinate system and quantization axis. With zero or very
small electric fields, the natural quantization axis is along the
magnetic-field direction $\boldsymbol{B}$. For even moderate electric
fields it becomes more convenient to define the quantization axis
along $\boldsymbol{E}$. In this study, we define our quantization
axis along the direction of the electric field but choose our
space-fixed $\hat z$ axis along $\boldsymbol{B}$. For convenience,
the laser propagates along our $\hat y$ axis and, hence, its polarization
$\boldsymbol{\hat{\epsilon}}$ lies in the $xz$ plane.  Finally,
$\theta$ is the angle between $\boldsymbol{\hat{\epsilon}}$ and
$\boldsymbol{B}$ and $\psi_m$ is the angle between $\boldsymbol{\hat{\epsilon}}$ and
$\boldsymbol{E}$.

We numerically solve for the eigenstates of Eq.~\ref{eq:ham} by
including basis states from $N=0$ up to $N_{\rm max}$.  This corresponds
to $(N_{\rm max}+1)^2(2I_a+1)(2I_b+1)$ basis states.  In the absence of
an electric field $N_{\rm max}=1$ is sufficient. For increasing electric
field, $N_{\rm max}$ must be increased as coupling to higher-lying angular
momentum states becomes more and more important.

\subsection{Pendular model for large electric fields}
\label{sec:simp}

For a finite electric field $H_\mathrm{E}$ quickly dominates, along
with $H_\mathrm{rot}$, over the other terms in Eq.~\ref{eq:ham}.
Hence, it is beneficial to investigate the energy structure of
$H_0=H_\mathrm{rot}+H_\mathrm{E}$ in the basis
$|N,m\rangle\equiv\phi_{v=0}(r) Y_{Nm}(\alpha\beta)|\Lambda\rangle$,
where the quantization axis is chosen along $\boldsymbol{E}$.  In
our model the dipole moment $\vec d$ in the DC Stark Hamiltonian
has spherical components $d_q= d_{v=0} C_{1q}(\alpha\beta)$, where
$d_{v=0}=\int_0^\infty dr\,\phi_{v=0}(r){\cal D}(r)\phi_{v=0}(r)$
is the $N$-independent vibrationally-averaged dipole moment  and
${\cal D}(r)$ is the $r$-dependent permanent electric dipole moment
of the X$^1\Sigma^+$ ground electronic state.

With our choice of quantization axis the electric field only couples
basis states $|N,m\rangle$ with the same $m$ and $N$ that
differ by one unit. In fact, for each $m$, $H_0$ is a
symmetric tridiagonal matrix with non-zero matrix elements $\langle
N,m|H_0|N,m\rangle = N(N+1)B_v $ and
\begin{equation}
  \langle N,m|H_0|N\!+\!1,m\rangle 
     = \frac{m^2-(N+1)^2}{\sqrt{(2N+1)(2N+3)}}d_{v=0}E \,.
\end{equation}
Its pendular eigenstates $|\lambda,m\rangle$ with $\lambda=0,1,\dots$
and corresponding eigenvalues ${\cal E}_{\lambda,m}$ have been
extensively studied in the context of the molecular orientation and
alignment \cite{HHughes1947,JRost1992,HLoesch1995} and are obtained
through numerical diagonalization.  At zero electric field strength
$\lambda=N$. For increasing field strengths  eigenstates of $H_0$
with the same $\lambda$ but different $|m|$ separate away from
each other, leaving a double degeneracy for states with $m \ne
0$.

The polarizability of  pendular states is determined from the derivative of 
eigenenergies of $H_0+H_{\rm pol}$ with respect to the laser intensity. For 
optical laser photons with an energy that is orders of magnitude larger than $B_v$
(and even vibrational spacings), the $N$-independent polarizabilities
$\alpha_{\parallel,\perp}(\omega)=\int_0^\infty dr \phi_{v=0}(r)
\alpha_{\parallel,\perp}(r;\omega) \phi_{v=0}(r)$, where
$\alpha_{\parallel,\perp}(r;\omega)$ is the radial electronic
polarizability.  Consequently, the eigenenergies of $H_0+H_{\rm pol}$ have a linear dependence
on $I_{\rm trap}$, the polarizability of 
pendular states is independent of  laser intensity, and  the 
so-called higher-order hyperpolarizabilities are zero.
In Ref.~\cite{Kotochigova2010} some of us showed the existence of a {\it magic} angle,
where the polarizability is insensitive to laser-intensity fluctuations. This occurs when  $C_{20}(\psi_m,0)=(3\cos^2\psi_m-1)/2=0$ 
or,  equivalently,  $\psi_m\approx54^\circ$.

\section{The $^{23}$N\MakeLowercase{a}$^{40}$K dimer}
\label{sec:NaK}

We can now investigate the energies and polarizabilities of
rotational-hyperfine states in the $v=0$  vibrational level of the
X$^1\Sigma^+$  electronic potential of $^{23}$Na$^{40}$K. Its
rotational constant is $B_{v=0}/h=2.8217297(10)$ GHz~\cite{Gerdes2008,WSebastian2016},
where $h$ is Planck's constant. The electric dipole moment
$d_{v=0}=1.07(2)ea_0$~\cite{Gerdes2008,WSebastian2016}, where $e$ is the electron
charge and $a_0$ is the Bohr radius.  One standard deviation
systematic and statistical uncertainties are given in parenthesis.
The nuclear spins are $3/2$
and $4$ for $^{23}$Na  and $^{40}$K, respectively.  The nuclear
electric quadrupole coupling constants  are  $(eqQ)_{\rm Na}/h=-0.187(35)$
MHz and $(eqQ)_{\rm K}/h=0.899(35)$ MHz~\cite{WSebastian2016}.  The two nuclear gyromagnetic ratios
are $g_{\rm Na}=1.477388(1)$ and $g_{\rm K}=-0.32406(6)$~\cite{EArimondo1977}.
The  frequency-dependent dynamic parallel and perpendicular
polarizabilities have been computed by us. A brief account of our
method  as well as numerical values are given in Sec.~\ref{sec:pol}.

A word on  energy scales is already in order. The hyperfine and
Zeeman interactions  as well as  $H_\mathrm{pol}$ have energies (in
units of the Planck constant) well below the MHz range as long as
the magnetic field strength is below $0.1$ T  and the laser intensity
is no larger than $10^4$ W$/$cm$^2$. These energy scales are much
smaller than $B_v$ as well as DC Stark shifts induced by reasonable
electric fields.  In order to limit our parameter space the magnetic
field $\boldsymbol{B}=B_z\hat z$ with $B_z=8.57$ mT throughout. This value was used  by Park {\it
et al.}~\cite{JPark2015}, who  formed weakly-bound Feshbach  molecules
at this field before performing a two-photon transition to create
ground-state molecules. If not otherwise specified, 
the laser intensity $I_\mathrm{trap}=2.35$ kW$/$cm$^2$,  typical
for ultracold experiments.

There are $36$ hyperfine states in each rotational manifold
$|N,m\rangle$. For electric fields up to  $10$ kV/cm, rotational
hyperfine states with $N$ up to $N_{\rm max}=5$ are incorporated in our numerical
calculations.

\subsection{Parallel and perpendicular polarizabilities}
\label{sec:pol}

\begin{figure}
 \includegraphics[width=1\columnwidth,trim=0 8 0 40,clip]{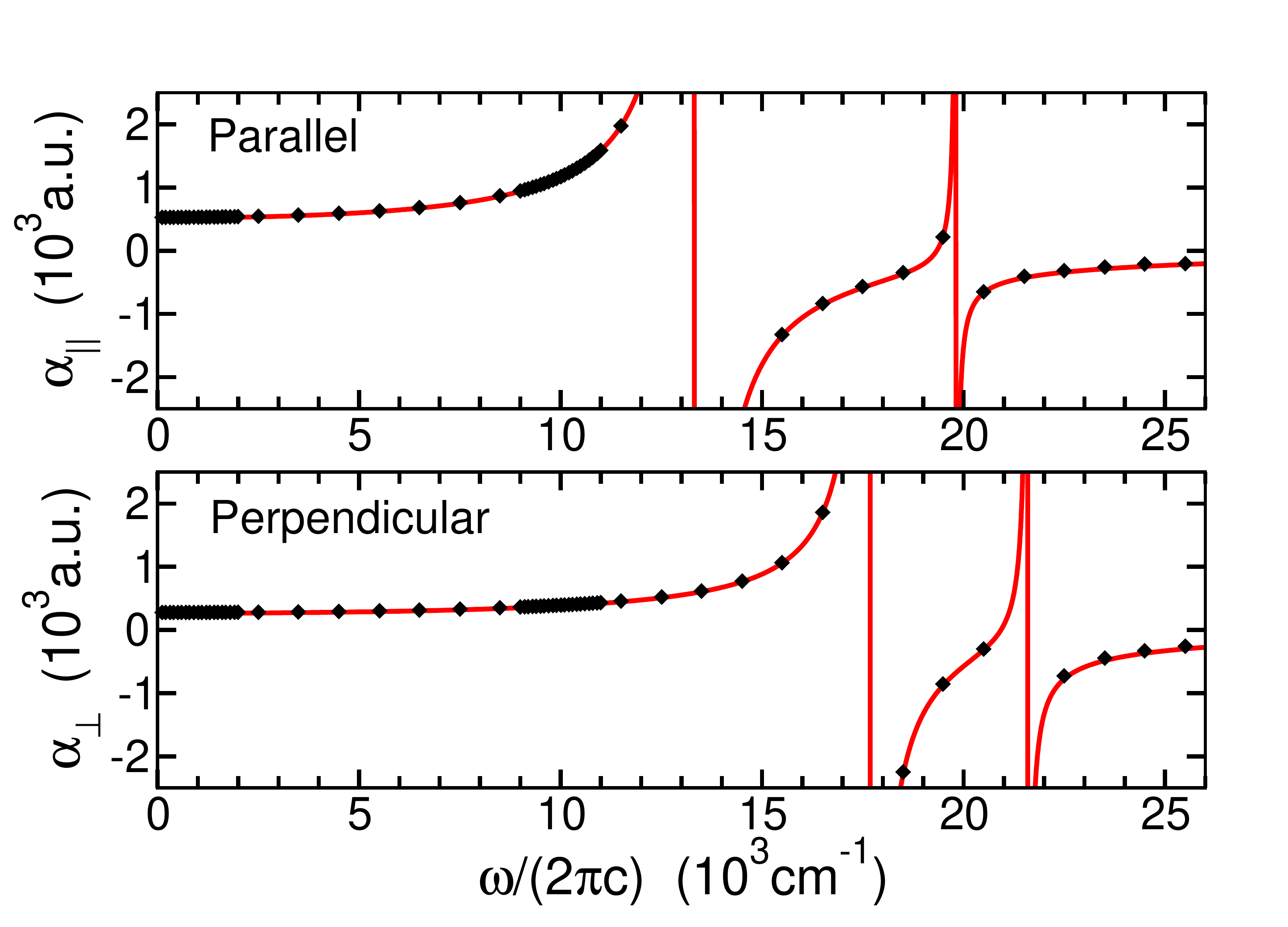}
 \caption{Dynamic parallel (top panel) and perpendicular (bottom
 panel) polarizabilities in atomic units,  a.u.=$(ea_0)^2/E_{\rm h}$, at
 the equilibrium separation   of  the ground $X^1\Sigma^+$ electronic
 state of NaK as  functions of laser frequency $\omega$. Black
 markers are our computed  data points, while the solid red curve
 corresponds to a fit to this data as described in the text.   The
 data is mainly localized near zero frequency and $\omega/(2\pi
 c)\approx 10\,000$ cm$^{-1}$. The latter corresponds to laser
 frequencies typically used for  trapping of ultra-cold molecules.
}
 \label{fig:alpha}
\end{figure}

The dynamic parallel and perpendicular radial electronic polarizabilities
$\alpha_{\parallel}(r;\omega)$ and $\alpha_{\perp}(r;\omega)$   can
be expressed in terms of a sum over all excited $^1\Sigma^+$  and
$^1\Pi$ electronic potentials, respectively.  We have calculated
these potentials and dynamic polarizabilities of NaK using the CCSD
method of the \textit{ab-initio} non-relativistic
electronic structure package CFOUR~\cite{cfour}.
Relativistic corrections are small for the relatively light Na and
K atoms. The def2-QZVPP basis sets of Ref.~\cite{Weigend2005} are
used for both atoms and include polarization functions. The specific
contraction of primitive basis functions are (20\textit{s} 12\textit{p}
4\textit{d} 2\textit{f})/[9\textit{s} 5\textit{p} 4\textit{d}
2\textit{f}] for Na and (24\textit{s} 18\textit{p} 4\textit{d}
3\textit{f})/[11\textit{s} 6\textit{p} 4\textit{d} 3\textit{f}] for
K. The computation is made tractable by only correlating valence
electrons and  electrons from the outer-most closed shell for each
atom.

Figure~\ref{fig:alpha}  shows the radial electronic
$\alpha_{\parallel}(r_e;\omega)$ and $\alpha_{\perp}(r_e;\omega)$
computed at the equilibrium separation $r_e=6.59a_0$ of the ground
X$^{1}\Sigma^+$ state as functions of laser frequency $\omega$. The
poles in the functions correspond to frequencies that are equal to
the energy difference  between an excited- and the ground-state
potential at $r_e$.  Our pole locations are consistent with the
potentials found in Ref.~\cite{Gerdes2008}.  Our calculated data
points from zero frequency up to  $\omega/(2\pi c)=30\,000$ cm$^{-1}$
are well described by
\begin{eqnarray}
\alpha_{\parallel}(r_e;\omega)&=&\frac{495.192}{1-(\nu/13322.2)^2}
                            +\frac{21.3802}{1-(\nu/19813.6)^2} ,\quad \\
\alpha_{\perp}(r_e;\omega)&=&\frac{228.684}{1-(\nu/17683.6)^2}
                            +\frac{34.6618}{1-(\nu/21595.1)^2} .\quad
\end{eqnarray}
in units of $(ea_0)^2/E_{\rm h}$  and $\nu=\omega/(2\pi c)$ in units
of cm$^{-1}$. Here, $E_{\rm h}$ is the Hartree energy and $c$ is
the speed of light. For $\nu < 21000$ cm$^{-1}$ deviations from the calculated
radial polarizabilities are no larger than  1\%  and 4\% for the parallel and perpendicular
polarizability, respectively, less than the uncertainty of the CCSD calculation.

The polarizability of the $v=0$ vibrational level is determined by
a vibrational average of $\alpha_{\parallel}(r;\omega)$ and
$\alpha_{\perp}(r;\omega)$. The radial $v=0$ wavefunction $\phi_{v=0}(r)$
is spatially localized around $r_e$ and the $r$-dependence of the radial
polarizabilities is small and, hence, to very good approximation
$\alpha_{\parallel,\perp}(\omega)$ are equal to the  corresponding
radial polarizability at $r_e$.  (Note also that the linear
dependence  of $\alpha_{\parallel,\perp}(r;\omega)$ near $r=r_e$ does not introduce corrections for 
nearly Gaussian $\phi_{v=0}(r)$.)

We find that the static (i.e.~zero frequency) polarizability for
the $N=0$, $v=0$ state
$[\alpha_{\parallel}(r_e;0)+2\alpha_{\perp}(r_e;0)]/3=348 (ea_0)^2/E_{\rm
h}$ in very good agreement with $351(ea_0)^2/E_{\rm h}$ from
Ref.~\cite{Deiglmayr2008}.  Furthermore, in most experimental
settings molecules are trapped using lasers with photon energies
that are well away from  electronic transitions.  Without loss of
generality, we choose a laser with a wavelength of 1064 nm as used
by or suggested in Refs.~\cite{JPark2015,MGempel2016}.  The two
$v=0$ polarizabilities are then $\alpha_\parallel=1013.4(ea_0)^2/E_{\rm
h}$ or  $\alpha_\parallel/h=4.749\times 10^{-5}$ MHz$/($W$/$cm$^2)$,
and $\alpha_\perp=361.46(ea_0)^2/E_{\rm h}$ or $\alpha_\perp/h=1.694\times
10^{-5}$ MHz$/($W$/$cm$^2)$.

\subsection{Energy levels and polarizabilities}

\begin{figure}
 \includegraphics[scale=0.33,trim=10 15 0 10,clip]{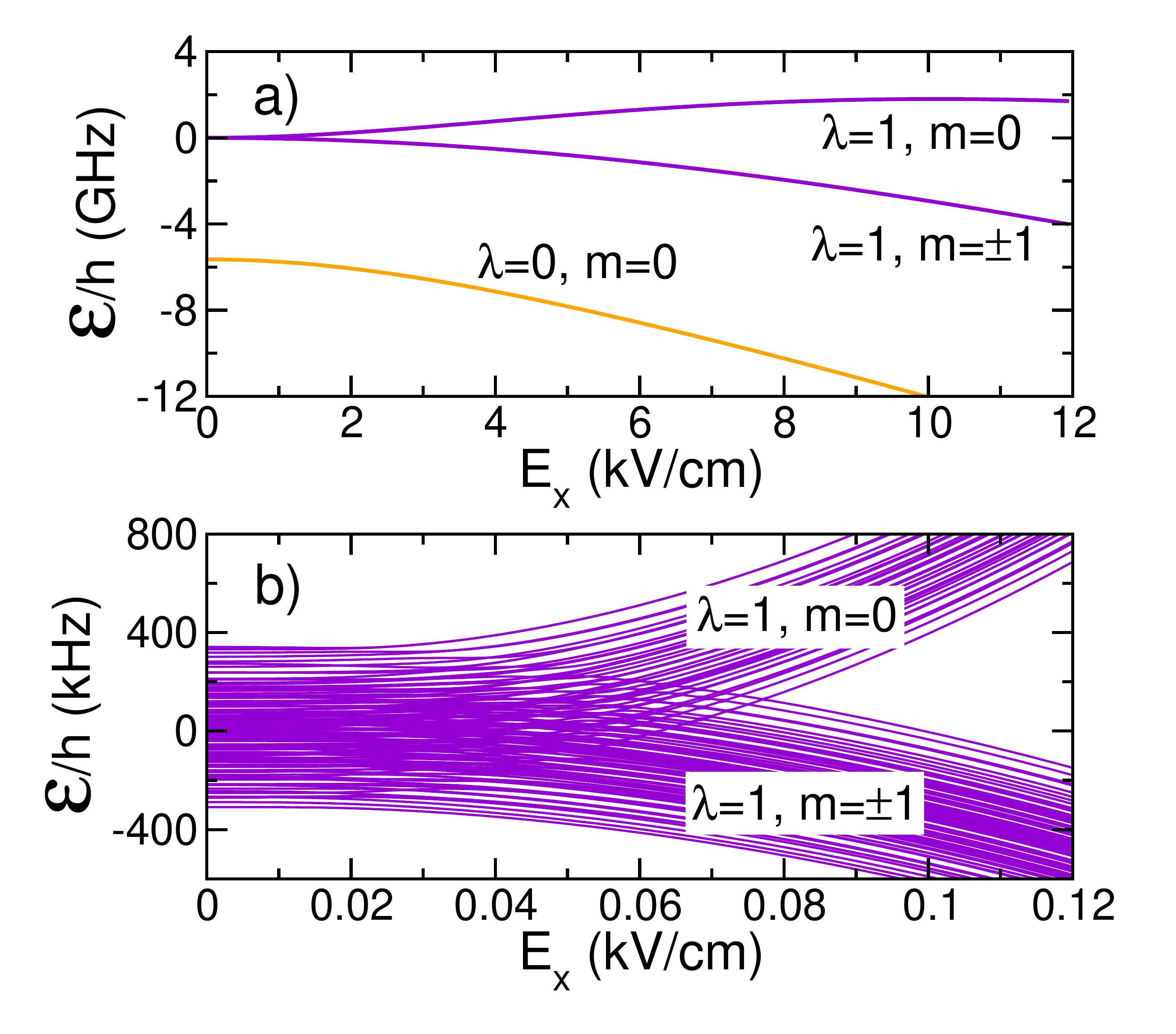}
\caption{Eigenenergies of the lowest rotational-hyperfine states
of the $v=0$ vibrational level of the electronic ground-state of
$^{23}$Na$^{40}$K as a function of static electric field strength $E_x$
when no trapping laser is present and $B_z=8.57$ mT.  The electric field
$\boldsymbol{E}=E_x \hat x$
is directed along our $\hat x$ axis. Panels a) and b) show the same data
on two different energy and electric field scales. Approximate
labels $\lambda$ and $m$ valid for  large electric fields are
indicated.  The zero of energy is  at the hyperfine barycenter of
the $N=1$ rotational state when $E_x=0$.}
\label{fig:enrg}
\end{figure}

Figure \ref{fig:enrg} shows the lowest 144 rotational-hyperfine
eigenenergies of $v=0$ ground-state $^{23}$Na$^{40}$K as a function
of static electric field strength $E_x$ when no trapping laser is
present. These levels correspond to 36  states in the $N=0$ manifold
and 108 states in the $N=1$ manifold.  The electric field 
is directed along our $\hat x$ axis. Panel a) shows these
eigenenergies on the scale of the rotational splitting, while panel
b) shows a blowup focussing on the $N=1$ manifold for ``small''
electric fields. We observe that the DC Stark effect  dominates
over $H_{\rm hf}$ and $H_{\rm Z}$ for $E_x>0.1$ kV/cm and states
can then be labeled by the pendular labels $\lambda=0,1$ and $|m|$.
When the electric field is near zero,  pendular states of the same
$\lambda$ with different $|m|$ are mixed by $H_{\rm hf}$ and
$H_{\rm Z}$ and the $N=1$ or equivalently $\lambda=1$ manifold has
a complex level structure.

\begin{figure*}[t]
\includegraphics[scale=0.4,trim=0 20 0 10,clip]{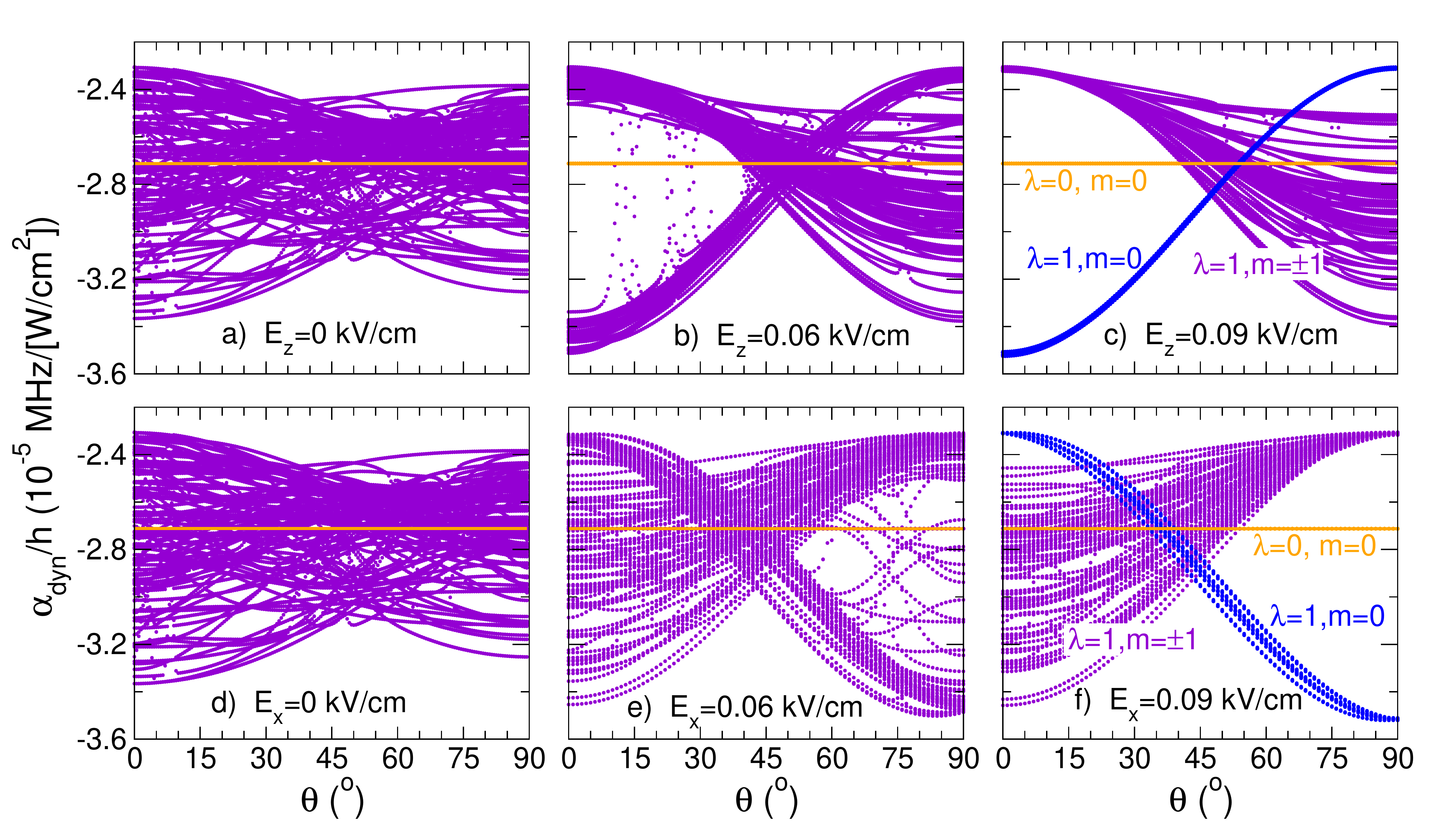}
 \caption{Dynamic polarizabilities of the lowest 144 eigen states
    of $^{23}$Na$^{40}$K as functions of the angle $\theta$ at small
    electric fields, $|\boldsymbol{E}|=0$ kV/cm, 0.06 kV/cm, and
    0.09 kV/cm.  Panels on the top and bottom row correspond to an
    electric field along the $\hat z$ and $\hat x$ axis, respectively.
    In panels a), b), d), and e) the orange line and purple markers
    correspond to $\lambda=0$ and $\lambda=1$ hyperfine states,
    respectively.  In panels c) and f) the  orange, blue, and purple
    lines and markers correspond to $|\lambda=0,m=0\rangle$,
    $|\lambda=1,m=0\rangle$, and $|\lambda=1,m=\pm1\rangle$ hyperfine
    states, respectively.  Panels a) and d) for zero electric field
    are identical.  The copy is only included for easy comparison
    with other panels.  We use $B_z=8.57$ mT, a laser wavelength of
    1064 nm, and $I_{\rm trap}=2.35$ kW/cm$^2$.}
\label{fig:smallExEz}
\end{figure*}

Figure \ref{fig:smallExEz} shows the dynamic polarizabilities
$\alpha_{{\rm dyn},i}$ of the $144$ hyperfine states in the $N=0$
and $1$ manifolds for   small electric field
strengths, ranging from 0 kV/cm to $0.09$ kV/cm, as functions of
$\theta$, the angle between the laser polarization and magnetic
field direction.  The laser has a wavelength of 1064 nm and $I_{\rm
trap}=2.35$ kW/cm$^2$.  The magnetic field $B_z=8.57$ mT and
the electric field is  applied along either the $\hat z$  or $\hat x$ direction.

At zero applied electric field, $\alpha_{{\rm dyn},i}$ of the
36 hyperfine states in the $N=0$ manifold are independent of $\theta$.
In fact, the total polarizability is  given
by~\cite{BNeyenhuis2012,KBonin1997} 
\begin{equation}
\alpha_{|N=0,m=0\rangle}(\omega)
     =(\alpha_{\parallel}(\omega)+2\alpha_{\perp}(\omega))/3 \,.
\end{equation}
On the other hand, the polarizability of states in  the $N=1$ manifold behave almost chaotically 
and is a consequence of strong mixing between the three $m$.  Thus,
a small fluctuation in the direction of the polarization will greatly
change the trapping potentials for these latter states. Also, each
of the corresponding eigenvectors changes drastically with a change
in $\theta$, making it difficult to focus on one eigenstate when
the directions of  external fields  change with respect to each
other during an experiment. This {\it non-adiabatic} admixing  is due
to the fact that the magnetic field is relatively small, and the
splitting between states that have similar hyperfine character but
different $m$ are comparable to $H_{\rm pol}$.

When an electric field is applied, the polarizability of $N=0$ or
$\lambda=0$ hyperfine states remain independent of $\theta$. The
polarizability of the $\lambda=1$ hyperfine states gradually group,
where the polarizability of  eigenstates dominated by  $m = 0$
character  start to coalesce into a single line on the scale of the
panels in Fig.~\ref{fig:smallExEz}.  The polarizability of $m=\pm1$
states also simplifies but remains a fairly complex for $\theta$
close to zero or 90 degrees.  This transition in behavior coincides
with the separation of eigenstate energies for states $|\lambda=1,
m=0\rangle$ from those with $|\lambda=1, m=\pm1\rangle$, as depicted by
Fig.~\ref{fig:enrg}b).

A comparison of the polarizability for electric fields along the
$\hat z$ and $\hat x$ direction and strengths larger than $0.06$ kV/cm shows
that the natural quantization axis is along electric field direction.
One manifestation is that the $\alpha_{\mathrm{dyn},i}$ for a field along
the $\hat x$ and $\hat z$ axis resemble each other when $\theta$ is replaced
by $90^\circ-\theta$.  I.e. for large fields the angular dependence
of $\alpha_{\mathrm{dyn},i}$ only depends on the angle between the laser
polarization and electric field. On the other hand the reflection
symmetry is not exact.  The grouping of the lines of $\alpha_{\mathrm{dyn},i}$
is not the same for the same $|\boldsymbol{E}|$, due to the remaining competition
between the Zeeman and DC Stark Hamiltonians.  For example, the 36
$\alpha_{\mathrm{dyn},i}$ of states in the $|\lambda=1,m=0\rangle$
manifold in Fig.~\ref{fig:smallExEz}f) are more spread out than
those in Fig.~\ref{fig:smallExEz}c).

\subsection{Single- and double-magic conditions}

\begin{figure*}
 \includegraphics[width=\textwidth,trim=0 0 0 0,clip]{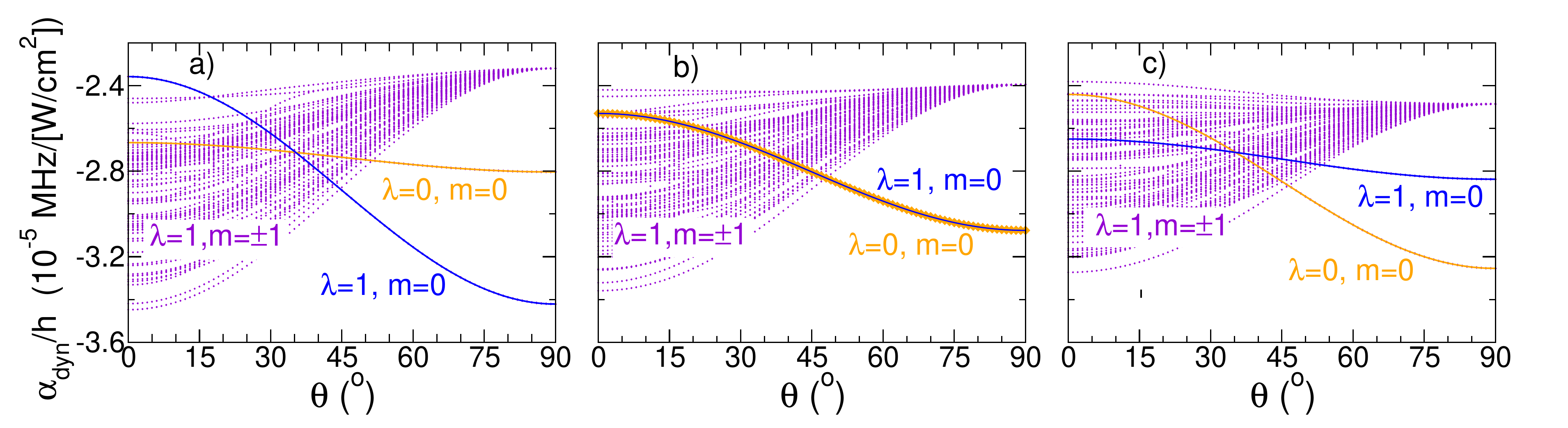}
  \caption{Dynamic polarizabilities of the lowest 144 rotational-hyperfine
     states of the ground vibrational level of $^{23}$Na$^{40}$K
     as functions of the angle $\theta$ for three strong electric
     fields $\boldsymbol{E}=E_x\hat x$ with $E_x=2.0$ kV/cm, $5.265$ kV/cm, and
     $8.0$ kV/cm in panel a), b), and c), respectively.  Orange,
     blue, and purple lines and markers correspond to hyperfine
     states in the $|\lambda=0,m=0\rangle$, $|\lambda=1,m=0\rangle$,
     and $|\lambda=1,m=\pm1\rangle$ manifolds, respectively.  We use $B_z=8.57$ mT, a laser wavelength of
    1064 nm, and $I_{\rm trap}=2.35$ kW/cm$^2$.}
\label{fig:magic} 
\end{figure*}

\begin{figure}
 \includegraphics[width=\columnwidth,trim=0 10 10 60,clip]{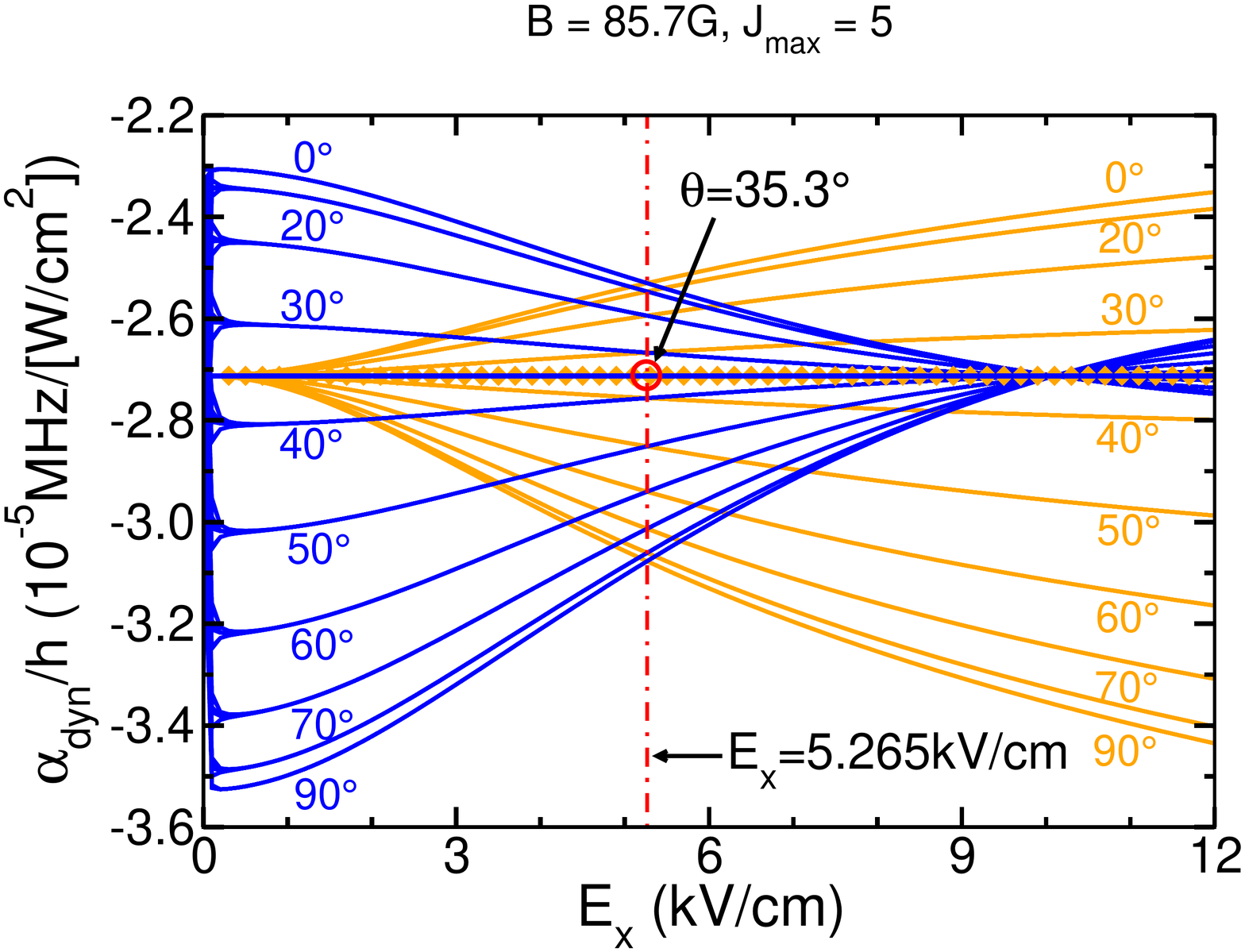}
 \caption{Dynamic polarizabilities of $m=0$ rotational-hyperfine
    states of the ground vibrational level of $^{23}$Na$^{40}$K as
    functions of the strength of an applied electric field for
    the eleven angles 
    $\theta=0^\circ,10^\circ,\cdots,80^\circ, 90^\circ$ and
    $35.3^\circ$. The electric field is directed along the $\hat x$ axis.
    Orange dashed and blue solid lines are polarizabilities for hyperfine states of
    the $|\lambda=0,m=0\rangle$ and $|\lambda=1,m=0\rangle$ manifold, respectively.    For
    $\theta=35.3^\circ$ the polarizability for the two manifolds
    is the same and independent of $E_x$.  At the red circle where
    $\theta=35.3^\circ$ and $E_x=5.265$ kV/cm our double {\it magic}
    condition holds. We use $B_z=8.57$ mT, a laser wavelength of
    1064 nm, and $I_{\rm trap}=2.35$ kW/cm$^2$.}
 \label{fig:AngEx}
\end{figure}

A careful study of Figs.~\ref{fig:smallExEz}c and \ref{fig:smallExEz}f
shows that {\it magic} conditions are  starting to occur. With the
electric field along the $\hat z$ axis and $E_z=0.09$ kV/cm the
polarizabilities $\alpha_\mathrm{dyn}$ of hyperfine states in the
$|\lambda=0,m=0\rangle$ and  $|\lambda=1,m=0\rangle$ manifolds are
almost the same near $\theta=54.7^\circ$ (or equivalently
near $\psi_m=54.7^\circ$).  This occurs
regardless of the hyperfine state in either manifold.  Moreover, for an electric field
along the $\hat x$ axis  {\it magic} conditions occur for 
$\theta\approx 90^\circ-54.7^\circ\approx35.3^\circ$.

We study this coalescence of the polarizabilities in more detail for
much larger electric field strengths and locate a case of double
{\it magic} conditions.  Figure~\ref{fig:magic} shows the polarizability
of states in the $\lambda=0$ and 1 manifolds for electric fields $E_x=2.0$ kV/cm, 
$5.265$ kV/cm, and $8.0$ kV/cm along the $\hat x$ axis. The polarizability
of the $|\lambda=0,1, m=0\rangle$ hyperfine states have now fully collapsed into one of
two $\theta$-dependent curves.  In fact, these $m=0$ polarizabilities are equal to better than 0.01 \% 
for both $\lambda=0$ and 1 hyperfine manifolds.
For the smallest and largest of the three strong electric fields the $|\lambda=0,m=0\rangle$
and $|\lambda=1,m=0\rangle$ curves cross at the {\it magic} angle
$54.7^\circ$. Crucially, for the {\it magic} intermediate electric
field strength of $5.265$ kV/cm, shown in Fig.~\ref{fig:magic}b), the
polarizabilities of all hyperfine state of the $|\lambda=0,m=0\rangle$
and $|\lambda=1,m=0\rangle$  manifolds coincide throughout the {\it
entire} range of $\theta$.  In fact, this {\it magic} electric field
strength exists regardless of field direction.

The dynamic polarizability of the $|\lambda=1,m=\pm1\rangle$ hyperfine states remains
very state dependent regardless of the electric field strength.  Here, the
electric field does not lift the degeneracy of $m=\pm1$ states, even though
the $N$-state mixture changes  with $|\boldsymbol{E}|$. The hyperfine
and Zeeman interactions then lead to ever changing couplings and dynanic polarizabilities.

Figure~\ref{fig:AngEx} shows the dynamic polarizabilities of
the $m=0$ hyperfine states of $^{23}$Na$^{40}$K but now
as functions of electric field strength for eleven angles $\theta$. The field is directed
along the $\hat x$ axis. We
see that for all angles the $\lambda=0$ and 1 polarizabilities cross at $E_x=5.265$ kV/cm.
At the special angle of $\theta=35.3^\circ$ these polarizabilities are the same for  {\it any}
electric field  $|\boldsymbol{E}|>0.25$ kV/cm. The red circle on this line corresponds to 
$|\boldsymbol{E}|=5.265$ kV/cm and a double {\it magic} condition where both the angular 
and electric field {\it magic}  conditions are met. In fact,  this double {\it magic} condition 
exists regardless of the field direction. It occurs when the angle between the direction 
of the laser polarization and the direction of the electric field 
is $\psi_m=54.7^\circ$ and $|\boldsymbol{E}|=5.265$ kV/cm.  Under normal 
fluctuations of experimental conditions, this double {\it magic} 
condition  provides extra stability for the matching of trapping 
potentials of  hyperfine states in the $|\lambda=0,m=0\rangle$ 
and $|\lambda=1,m=0\rangle$ manifolds. The same concept does  
apply to other ultracold dipolar species. The value for the {\it magic} $|\boldsymbol{E}|$ will be different, of course, and
is determined by the rotational constant, the permanent dipole moment, as well as the radial electronic polarizability.

\subsection{Sensitivity to the laser intensity}
\begin{figure}
 \includegraphics[scale=0.3,trim=0 0 0 10,clip]{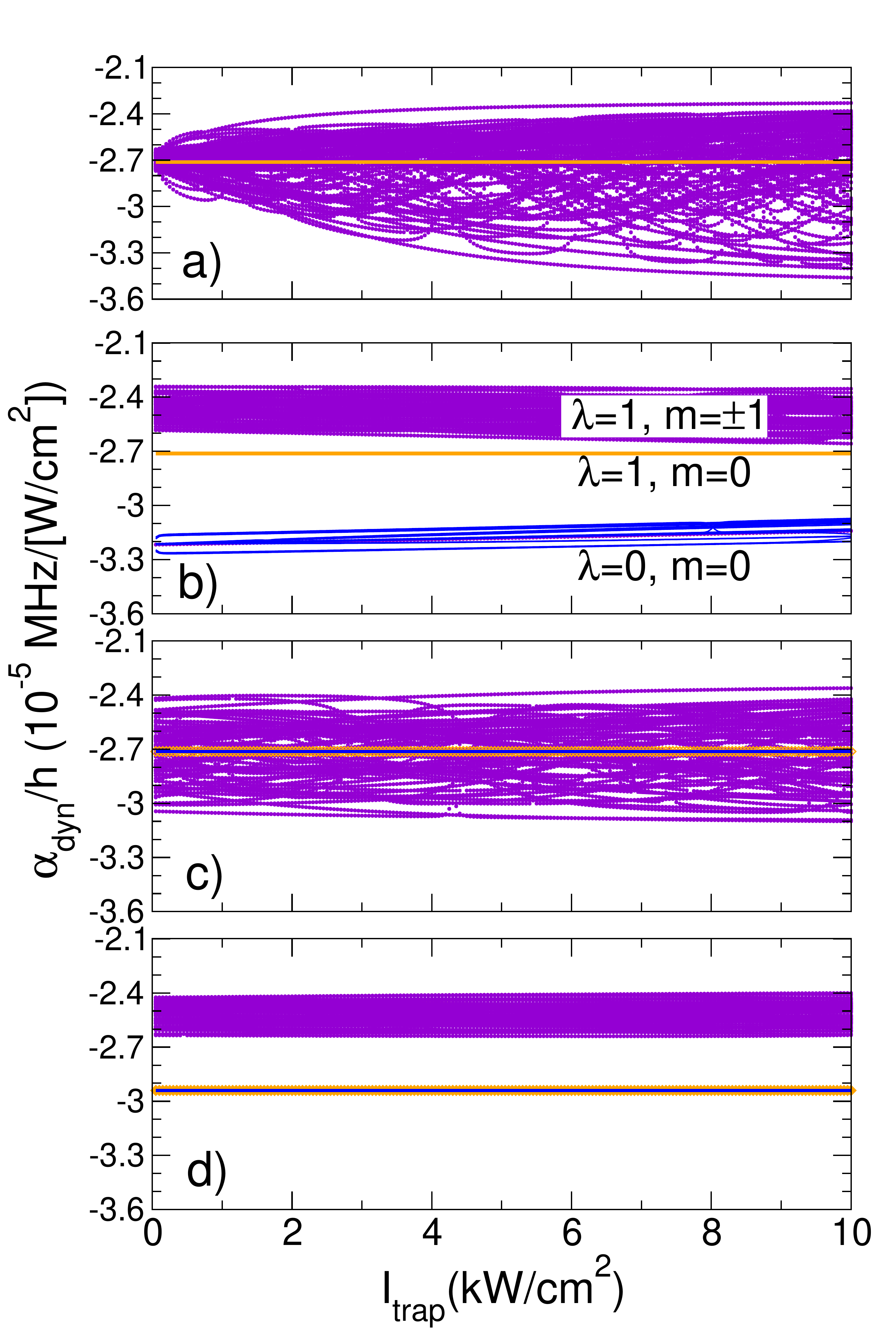}
 \caption{Dynamic polarizabilities of the lowest 144 eigenstates of the vibrational ground state of $^{23}$Na$^{40}$K as functions
     of the trapping laser intensity $I_{\mathrm{trap}}$ at four representative
     pairs $(E_x,\theta)=(0.0,60^\circ$),  $(0.09, 60^\circ)$, $(2.0, 35.3^\circ)$ and $(5.265, 60^\circ)$ in panels a), b), c) and d), respectively. The electric field is in units of kV/cm and pointed along the $\hat x$ axis.
     In all panels the orange dashed line corresponds to  $\alpha_{\rm dyn}$ of the 36 $\lambda=0,m=0$
     hyperfine states.  In panel a)  the purple dots correspond to the 108 $\lambda=1$
     hyperfine states with mixed $m$ character. In panels b), c), and d)
     the blue  and purple dots are eigenstates
     dominated by  $|\lambda=1,m=0\rangle$  and (mixed) $|\lambda=1,m=\pm 1\rangle$ pendular functions.
     In panels c) and d) the polarizibilities of the  $m=0$ eigenstates are indistinguishable. Both correspond 
     to  {\it magic}  conditions for $^{23}$Na$^{40}$K.
     We use $B_z=8.57$ mT and a laser wavelength of 1064 nm.
     }
\label{fig:hyper}
\end{figure}

A second benefit of applying a strong electric field is the negligible   
dependence of the dynamic polarizabilities of  $m=0$ hyperfine states on
laser intensity, consistent with the prediction of the pendular model in Sec.~\ref{sec:simp}. 
In Figs.~\ref{fig:hyper}, the polarizabilities of the 
144 states in the $\lambda=0$ and $1$ manifolds are plotted as functions of 
laser intensity  for four representative pairs $(E_x,\theta)$. When the electric 
field is small or zero, i.e. $E_x\ll 0.1$ kV/cm, the 108 eigen states in the $\lambda=1$ manifold
are mixed with respect $m=0,\pm1$ and are sensitive to fluctuations in the experimental conditions, including that of the
trapping laser intensity. Figure~\ref{fig:hyper}a shows that the
dependence of the polarizabilities of these states on $I_\mathrm{trap}$
is complicated. Even if the polarizabilities are 
matched for two states at a certain intensity,  small fluctuations
 introduce a mismatch of the polarizabilities and thus the trapping
potentials. As the electric field is increased to $0.1$ kV/cm  some states start
to be dominated by $m=0$ character and separated from the others.
Their polarizabilities  group, as shown in 
Fig.~\ref{fig:hyper}b). As the strength of the electric field is
further increased the polarizabilities of states dominated by $|\lambda=1,m=0\rangle$ character
coincide and become independent of the intensity. This is demonstrated at the {\it magic} angle and the
{\it magic} electric field  in Figs.~\ref{fig:hyper}c)
and \ref{fig:hyper}d), respectively, where the polarizabilities of the
$|\lambda=1,m=0\rangle$ states also equal that of  $|\lambda=0,m=0\rangle$ states. 
The polarizabilities of $|\lambda=1,m=\pm1\rangle$ states still remain sensitive to
laser intensity fluctuations.

We  quantify the dependence of the polarizabilities on the intensity with
the difference in the hyperpolarizabilites between the 
$|\lambda=0,m=0\rangle$ and $|\lambda=1,m=0\rangle$ states, where
the hyperpolarizability of state $i$ is defined as 
$\beta_i=d\alpha_\mathrm{dyn,i}/dI_\mathrm{trap}$, where
the electronic hyperpolarizability of $\alpha_{||}(r;\omega)$ and $\alpha_{\perp}(r;\omega)$
can be safely neglected. At the double {\it magic} condition, the difference
between the two hyperpolarizabilities is $\sim$0.03 Hz/[kW/cm$^2$]$^2$.
This implies that a change of the trapping laser intensity of 
1 kW/cm$^2$ will result in the change of the polarizability by 
about one part in a million.  Hence, the intensity dependence of the total 
polarizability is insignificant and can be neglected.

\section{Summary}
\label{sec:sum}

We have shown that a strong  electric field  can effectively decouple 
rotational and nuclear degrees of freedom of ultracold  polar di-atomic molecules held
in optical lattices. This decoupling can be used  to prepare pairs of rotational-hyperfine 
states that exhibit fluctuation-insensitive {\it magic} trapping conditions. The two states
then have the same dynamic polarizability.
These  {\it magic}  conditions can be either single or double in nature by giving stability
against one or two distinct types of fluctuations.

Our theoretical predictions are based on a quantitative Hamiltonian
for ro-vibrational, hyperfine states of $^1\Sigma^+$ molecules in the
presence of various external electro-magnetic fields. These include
a magnetic field, the static electric field, and  trapping laser fields.
Among them, the electric field is especially 
useful in simplifying the theory for states dominated by the rotational projection
quantum number $m=0$  and, thereby, leads to our hyperfine-state insensitive {\it magic} 
trapping conditions.

We studied the electronic ground-state $^{23}$Na$^{40}$K molecule as an 
important test case and used its newly calculated parallel and perpendicular electronic
dynamic polarizabilities.  For strong electric fields a {\it magic} angle of $\psi_m=54.7^\circ$
was found, which protects against fluctuations in the angle between the laser polarization and electric field
for the 72 hyperfine states with $|\lambda=0,m=0\rangle$ and $|\lambda=1,m=0\rangle$ character. 
We also predicted a double {\it magic} condition at an electric field strength of $5.26(15)$ kV/cm 
and angle $\phi_m=54.7^\circ$. It provides  stability against electric field strength fluctuations. 
For these $m=0$ hyperfine states the laser-intensity dependence 
of the dynamic polarizabilities is shown to be insignificant. 
The one-standard deviation uncertainty of the {\it magic} electric field is due to the
combined uncertainty of the permanent electric dipole moment and the parallel and perpendicular
electronic polarizabilities. For $|m|=1$ states the dynamic
polarizability shows complex angle, field strength and intensity dependence due the near energy
degeneracy of these states independent of electric field strength.
We gave physical intuition based on the simplified, pendular theory.

\begin{acknowledgments}
This work is supported by grants from the United States Army Research Office, MURI Nos.
W911NF-14-1-0378 and W911NF-12-1-0476 and the United States National Science Foundation No. PHY-1619788.
\end{acknowledgments} 
\bibliography{NaK}

\end{document}